\documentclass[fleqn,twoside]{article}
\usepackage{espcrc2}
\usepackage{graphicx}
\makeatletter
\newcommand\nameaddress[1]{{\addtocounter{address}\m@ne
                            \expandafter\xdef
                               \csname address.#1\endcsname
                               {\number\value{address}}%
                            \addtocounter{address}\@ne}}
\newcommand\useaddress[1]{{\edef\doit{\noexpand\addressmark
                                      \noexpand\setcounter{address}%
                                      {\number\value{address}}}%
                           \setcounter{address}{\csname address.#1\endcsname
                                                }%
                           \expandafter}\doit}
\newcommand\anotheraddress[1]{{\let\orig@makeadmark=\@makeadmark
                               \def\@makeadmark##1{\orig@makeadmark
                                                   {\negthinspace,##1}}%
                               \address{#1}}}
\newcommand\slashnext[1]{\mathpalette{\bgroup\let\style=}
                                     {\setbox0=\hbox{$\style #1$}%
                                      \setbox2=\hbox to\wd0{\hss$\style/$\hss}%
                                      \hbox to 0pt{\box2\hss}\box0\egroup}}
\@ifundefined{inlinecite}{}

\newcommand{\Sign}{\mbox{Sign}}

\newcommand{\e}{{\mathrm{e}}}

\makeatother
\begin{document}
\title{\begin{flushright}\normalsize
            DUKE-TH-00-208 \\
       \end{flushright}
       Critical Behavior of a Chiral Condensate with a
       Meron Cluster Algorithm} 
\author{Shailesh Chandrasekharan and James C. Osborn,\thinspace
        \address{Department of Physics, Duke University, 
        Durham, NC 27708-0305, USA}%
        }

\begin{abstract}
A new meron cluster algorithm is constructed to study the 
finite temperature critical behavior of the chiral condensate 
in a $(3+1)$ dimensional model of interacting staggered fermions. 
Using finite size scaling analysis the infinite volume condensate 
is shown to be consistent with the behavior of the form 
$(T_c-T)^{0.314(7)}$ for temperatures less than the critical
temperature and $m^{1/4.87(10)}$ at the critical temperature
confirming that the critical behavior belongs to the 3-d Ising 
universality class within one to two sigma deviation. The new method,
along with improvements in the implementation of the algorithm,
allows the determination of the critical temperature $T_c$ more 
accurately than was possible in a previous study. 
\end{abstract}
\maketitle

\section{Motivation}

The construction of Monte Carlo algorithms to solve problems in many
body quantum mechanics involving fermions is notoriously difficult.
This difficulty is reflected in our inability to perform precise
quantitative calculations in strongly interacting fermionic 
models which are necessary to understand a variety of phenomena 
including high temperature superconductivity and the physics of strongly 
interacting matter. The essential problem arises due to the Pauli 
principle which can produce negative Boltzmann weights when the quantum 
partition function is rewritten as a path integral in a convenient basis. 
As a result, the probability distribution that should be used for 
importance sampling is unclear.

The conventional approach to problems involving fermions is to integrate 
them out in favor of a fermion determinant. In cases where this determinant 
is positive it is often possible to use known sampling methods for a 
bosonic problem to devise an algorithm \cite{Dua85,Hor85,Whi88,Lus94}. 
Some of these methods are inexact since they involve discretization of a 
differential equation and require some care and study before they can be 
applied to a new problem. Others can suffer from truncation errors that 
approximate the original partition function. Worst of all, these algorithms 
suffer from the usual problems of critical slowing which makes it difficult 
to study phase transitions using them.

The study of phase transitions, especially in the context of fermionic
models, is of interest in a variety of fields. In condensed matter
physics strong correlations between electrons can lead to many interesting
critical effects like high $T_c$ superconductivity \cite{And87} and 
quantum phase transitions \cite{Son97}. In high energy physics, existence 
of new phases and fixed points have been predicted \cite{Kog88,Ros89,App96} 
which may lead to novel formulations of quantum field theories beyond 
perturbation theory. Additionally, exotic phases arise in dense nuclear matter
due to strong interactions among quarks \cite{Kri98}.

Since fermions acquire a screening mass on the order of the
temperature $T$, one expects that the finite temperature critical
behavior close to a second order phase transition in a $(d+1)$ dimensional
theory is governed by a $d$ dimensional low energy effective theory that 
is purely bosonic \cite{Pis84}. A few years ago, this conventional wisdom 
was questioned based on a large N calculation \cite{Kocl95} in a Gross-Neveu 
model. It was shown that the finite temperature phase transition
in the $(2+1)$ dimensional model reproduced mean field exponents instead 
of the expected 2-d Ising exponents. This claim was later backed by 
numerical evidence \cite{Kocr95} using the hybrid Monte Carlo algorithm. 
Although, the reason for the unexpected critical behavior was later 
attributed to the narrowness of the Ginsburg region in the large N 
limit \cite{Kog98}, the numerical evidence provided to substantiate the 
earlier claims remains disturbing and perhaps shows the inadequacy of the 
numerical methods used. Recently, the chiral transition in two flavor QCD 
with an additional four-fermion interaction was studied using the hybrid 
molecular dynamics algorithm which showed evidence for non-mean field critical 
exponents \cite{Kog00}. However, again the expectations based on simple
dimensional reduction were not observed, instead the data appeared to be
consistent with tricritical behavior. 

The inability to provide conclusive answers to questions related to
the critical behavior in fermionic theories is closely related to the
lack of efficient fermion algorithms.
The fermion cluster algorithms, recently proposed in \cite{Cha99.1,Cha99.2},
provide a novel approach to the problem. The new method, referred to as 
the {\sl meron cluster algorithm}, uses well known quantum cluster algorithm 
techniques \cite{Eve93} to solve the fermion sign problem completely.
The first applications of the meron algorithm emerged last year when
it was used to study the critical behavior in a relativistic system of 
interacting staggered fermions with a discrete chiral symmetry 
\cite{Cha99.3,Cox00}. The results indicated that the symmetry is broken by 
the ground state but is restored by thermal fluctuations at high 
temperatures. However, in order to avoid the complications that arise in 
the algorithm due to the addition of a mass term, the previous 
study focused on massless fermions. Since the chiral condensate vanishes 
in this case the scaling of the chiral susceptibility 
with the volume was used to find the critical temperature and the critical 
exponents $\nu$ and $\gamma$.

In this article a new meron algorithm is proposed and applied to study 
the staggered fermion model in the presence of the mass 
term\footnote{Preliminary results of this work was presented in
\cite{Cha99.ch}}. This makes it possible to study the critical behavior 
of the chiral condensate. The main
result of the article is that in $(3+1)$ dimensions the chiral condensate 
behaves like $\langle\bar\psi\psi\rangle = A (T_c-T)^\beta$ just below
$T_c$ with $\beta=0.314(7)$ and vanishes for higher temperatures. The 
corresponding exponent in the 3-d Ising model is known to be 
$0.32648(18)$. As a bonus the critical temperature is also determined 
more accurately than previous studies. Furthermore at $T_c$ the mass 
dependence of the condensate also behaves as expected, 
$\langle\bar\psi\psi\rangle = B m^{1/\delta}$, where $\delta=4.87(10)$ 
as compared to $4.7893(22)$ of the Ising model. Conventional fermionic 
algorithms have never been able to confirm the predictions of 
universality in a strongly interacting fermionic model with such 
precision. 

\section{The Model}

The Hamilton operator for staggered fermions hopping on a 3-d cubic 
spatial lattice with $V = L^3$ sites ($L$ even) and anti-periodic 
spatial boundary conditions, considered here, is given by
\begin{equation}
H = \sum_{x,i} h_{x,i} + m \sum_x s_x.
\label{Ham}
\end{equation}
The term 
\begin{equation}
h_{x,i} = \frac{\eta_{x,i}}{2}(c_x^\dagger c_{x+\hat i} + 
c_{x+\hat i}^\dagger c_x) + (n_x - \frac{1}{2})
(n_{x+\hat i} - \frac{1}{2}),
\end{equation}
couples the fermion operators at the nearest neighbor sites $x$ and 
$x+\hat i$ where $\hat i$ is a unit-vector in the positive $i$-direction 
and the mass term 
\begin{equation}
s_x =  (-1)^{x_1+x_2+x_3} (n_x-1/2),
\end{equation}
is a single site operator. The fermion creation and annihilation operators 
$c_x^\dagger$ and $c_x$ satisfy the canonical anti-commutation relations
and $n_x = c_x^\dagger c_x$ is the number operator. The phase factors 
$\eta_{x,1} = 1$, $\eta_{x,2} = (-1)^{x_1}$ and 
$\eta_{x,3} = (-1)^{x_1 + x_2}$ are well known in the staggered fermion
formulation \cite{Sus77}. 

This model was originally studied in the chiral limit ($m=0$) in 
\cite{Cha99.3}. In this limit the Hamilton operator is invariant under 
shifts in the $x_3$ direction. The mass term breaks this symmetry up to 
shifts by an even number of lattice units, thus breaking a $\mathbf{Z}_2$ 
symmetry which can be related to a subgroup of the well known chiral 
symmetry of relativistic massless fermions. This symmetry is broken 
spontaneously at zero temperatures, while thermal fluctuations restore it 
at some high temperatures \cite{Cha99.3}. In this article the critical
behavior near the second order transition is studied using the chiral 
condensate $\langle \bar\psi\psi\rangle \equiv -\sum_x \langle s_x \rangle/V$
using the algorithm presented in \cite{Cha99.ch} and briefly sketched here.

The construction of the path integral for the partition function is 
standard. First, the Hamilton operator is rewritten as 
$H = H_1 + H_2 + ... + H_6 + H_7$ with
\begin{equation}
H_i = \!\!\! \sum_{\stackrel{x = (x_1,x_2,x_3)}{x_i even}} \!\!\! h_{x,i}
\, \;\;\;\;
H_{i+3} = \!\!\! \sum_{\stackrel{x = (x_1,x_2,x_3)}{x_i odd}} \!\!\! h_{x,i}
\end{equation}
for $i=1,2,3$ and $H_7 = m \sum_x s_x$. The partition function is then
approximated by
\begin{equation}
\mbox{Tr} 
[\e^{- \epsilon H_1} \e^{- \epsilon H_7/6} \e^{- \epsilon H_2} 
\e^{- \epsilon H_7/6}... \e^{- \epsilon H_6}\e^{- \epsilon H_7/6}]^M
\label{partfn}
\end{equation}
where the inverse temperature has been divided into $M$ slices such that
$M\epsilon = 1/T$. At a fixed temperature, the above approximation becomes
exact in the limit $M\rightarrow\infty$ and $\epsilon \rightarrow 0$.
On the other hand for any fixed $M$, the approximation defines a
new theory with a phase structure and critical behavior that can be 
identical to the $M=\infty$ theory. For simplicity this article focuses 
on the theory with $M=4$.

\begin{figure}
\begin{center}
\vskip-0.8in
\includegraphics[width=0.45\textwidth]{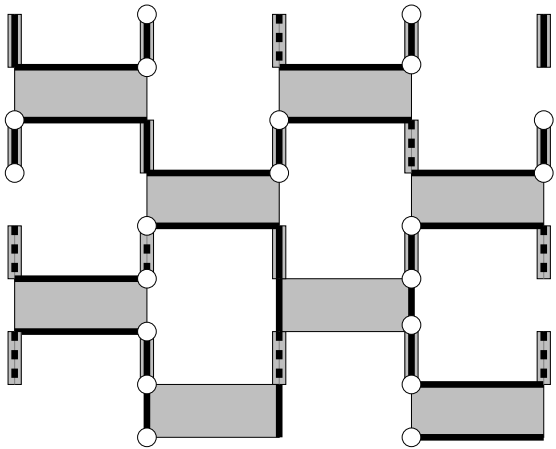}
\end{center}
\vskip-1.2in
\caption{ A configuration of fermion occupation numbers and bonds 
on a one dimensional periodic spatial lattice of size $L=4$. Each shaded
region corresponds to a transfer matrix element and has weights and
signs that are shown in figure \ref{elements}.}
\label{fconf}
\end{figure}

\section{Meron Cluster Algorithm}

\begin{figure}
\begin{center}
\vskip-0.8in
\includegraphics[width=0.45\textwidth]{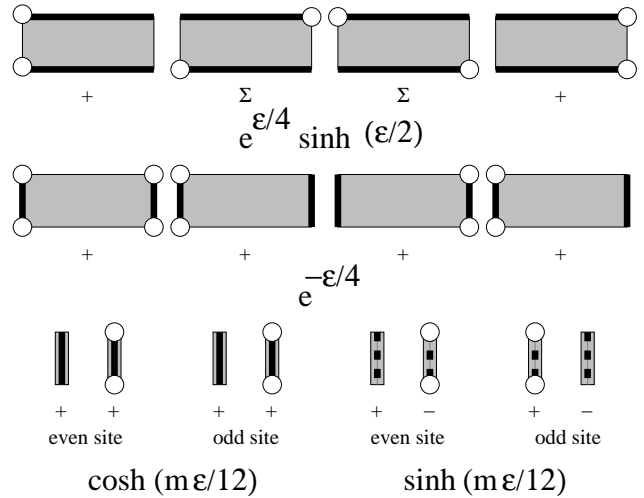}
\end{center}
\vskip-1.2in
\caption{ The magnitude and sign of the non-zero transfer matrix 
elements. The product of such weights over the entire lattice 
determine $W[n,b]$ and $\Sign[n,b]$. The signs $\Sigma$ get 
contributions from local staggered fermion phase factors and 
non-local factors that arise due to anti-commutation relations.}
\label{elements}
\end{figure}

In order to solve the model discussed in the previous section using
a meron cluster algorithm, the partition function should first be
written in terms of fermion occupation numbers $n$ and bond variables 
$b$ so that it can be represented by
\begin{equation}
Z = \sum_{[n,b]} \Sign[n,b] W[n,b].
\end{equation}
Configurations $[n,b]$ live on an $L^d\times 4dM$ lattice. A typical 
configuration in one space and one time dimension is shown in figure 
\ref{fconf}. The shaded regions represent either the nearest neighbor 
interactions that arise due to terms of the form $\exp(-\epsilon h_{x,i})$ or 
the single site interactions due to mass terms $\exp(-\epsilon m s_x/6)$. The 
various steps that are used to determine $W[n,b]$ and $\Sign[n,b]$ 
starting from the partition function (\ref{partfn}) have been discussed 
in \cite{Cha99.ch,Cha99.3,Wie93}. The final results can be represented
as a set of simple rules. $W[n,b]$ turns out to be a product of 
magnitudes of the transfer matrix elements and $\Sign[n,b]$ represents 
the product of their signs. The non-zero matrix elements are shown in 
figure \ref{elements} along with their weights. When compared to 
\cite{Cha99.3} the only difference is in the mass term. Assuming 
$m\;\geq\;0$ in (\ref{Ham}) leads to two types of single 
site interactions. The one with the solid bond is always positive,
and the one with the dotted bond is negative on filled even sites
and empty odd sites. This extra negative sign must be included in
$\Sign[n,b]$ along with sign $\Sigma$ that arises due to 
fermion hops, staggered fermion phase factors and anti-periodic
spatial boundary conditions as discussed in \cite{Cha99.3}.

Bonds connect lattice points into clusters. A flip of a 
cluster is defined as the change in the configuration of fermion 
occupation numbers at the sites belonging to the cluster, such that 
an occupied site is emptied and vice-versa. For a given model to be
solvable using a meron cluster algorithm, the weights $W[n,b]$ and
signs $\Sign[n,b]$ must satisfy three properties under cluster flips.
\begin{enumerate}
\item The weight of the configuration $W[n,b]$ must not change under
the flip of any cluster.
\item The change in $\Sign[n,b]$ due to a cluster flip must be independent
of the state of other clusters.
\item Starting from any configuration $[n,b]$, it must be possible to
reach a reference configuration $[n_{\rm ref},b]$ by flipping clusters 
such that $\Sign[n_{\rm ref},b]$ is 1.
\end{enumerate}
As has been discussed in \cite{Cha99.at}, the change in the sign of
the configuration due to a cluster flip depends on the topology
of the cluster. If we refer to the clusters whose flip changes the 
sign of the configuration as {\em merons}, then it is easy to classify 
a configuration $[n,b]$ based on the number of meron clusters it contains. 
Using the above three properties it is 
then easy to check that the partition function gets contributions only from
the zero meron sector, i.e.,
\begin{equation}
Z = \sum_{[n,b]} \overline{\Sign[n,b]} W[n,b].
\end{equation}
where $\overline{\Sign[n,b]} = \delta_{N,0}$ and $N$ denotes the number of
merons in the configuration. For the present model it is also easy to
show that the chiral condensate is given by
\begin{equation}
\langle \bar\psi \psi\rangle = \frac{1}{V Z} \sum_{[n,b]}\; 
{\rm Size}({\cal C}_{\rm meron})\;\delta_{N,1}\;W[n,b],
\end{equation}
which means that to measure the condensate one is interested only
in the zero and one meron number sectors.

  A typical sweep in a meron cluster algorithm consists of two 
steps exactly like any other cluster algorithm. Starting from a 
configuration the bonds are updated first based on the weights 
$W[n,b]$. After each local bond update the meron number of
the configuration can potentially change. In order not to generate 
more than the necessary number of merons every bond update is 
followed by a Metropolis decision. While measuring the chiral 
condensate for example one rejects all bond updates that generate
more than one meron. This requires one to reanalyze the topology
of clusters that are connected to the local bonds being updated.
A major improvement in the implementation allows this 
to be done in at most $\log({\rm Size}({\cal C}))$ steps \cite{Osb00}.
Once all the bonds are updated it is possible to update the occupation
numbers by flipping each cluster with a probability half.  This
algorithm produces the configuration $[n,b]$ with weight 
$ (\delta_{N,0} + \delta_{N,1}) W[n,b]$. Thus, the condensate 
can be calculated using
\begin{equation}
\langle \bar\psi \psi\rangle = 
\frac{\langle \;{\rm Size}({\cal C}_{\rm meron})\;\delta_{N,1}\rangle}
{V \langle \;\delta_{N,0}\;\rangle} \;,
\label{cceq}
\end{equation}
where $\langle \ldots \rangle$ refers to a simple average over the generated
configurations.

\section{Numerical Results}

The critical behavior of the model was studied through
the measurement of the chiral condensate at several values 
of the temperature around $T_c$ using the algorithm discussed
above. For each temperature simulations were performed
on different spatial lattice sizes and at various masses.  
Each simulation produced $10^6$ configurations, except for 
the largest lattices of sizes ranging from $32^3$ up to 
$48^3$ which contained only $10^5$ configurations. All runs 
included at least ten thousand thermalization sweeps in addition 
to the above measurements. The autocorrelation times typically
ranged from two to five sweeps. However, errors were evaluated 
from fluctuations in the averages of data over blocks of 1000 
configurations each. For future reference we give the values 
of the chiral condensate on a $32^3$ lattice obtained at $m=0.001$ 
and various temperatures in the table below.
\begin{center}
\begin{tabular}{|c|c||c|c|}
\hline
T & $\langle\bar\psi\psi\rangle$ & T & $\langle\bar\psi\psi\rangle$ \\
\hline
1.0000 & 0.270(20) & 1.0471 & 0.1478(61) \\ 
1.0152 & 0.228(16) & 1.0638 & 0.0566(17) \\
1.0309 & 0.205(12) & 1.0870 & 0.0175(04) \\
\hline
\end{tabular}
\label{data}
\end{center}

In order to confirm the spontaneous breaking of the 
$\mathbf{Z}_2$ chiral symmetry the condensate needs to be evaluated 
in the infinite volume limit followed by the chiral limit. 
This can be done precisely by a finite size scaling analysis.
In the broken phase the theory undergoes a first order phase 
transition as a function of the mass at $m=0$ where the 
condensate exhibits a jump. In a large but finite volume this 
discontinuity is smoothened out to an analytic curve whose 
functional form is given by \cite{PF83}
\begin{eqnarray}
\langle \bar\psi\psi \rangle = \Sigma_0 \tanh(m V \Sigma_0 /T) +
 \chi_0 \; m \; ,
\label{fss1}
\end{eqnarray}
which is valid when 
\begin{equation}
m \ll \Sigma_0 / \chi_0 \;.
\label{mcond}
\end{equation}
By fitting the available data at each simulation temperature to 
the formula (\ref{fss1}) it is possible to extract $\Sigma_0$,
the desired limiting value of the condensate. The minimum 
volume necessary for the formula to work can be determined by 
systematically removing the smallest volume data from the fit, as 
required to obtain a $\chi^2/DOF$ of about one. Finally, it 
is important to check if $\Sigma_0$ and $\chi_0$ obtained 
through the fit and the masses used are consistent with the 
condition (\ref{mcond}). 

\begin{figure}
\begin{center}
\includegraphics[width=0.45\textwidth]{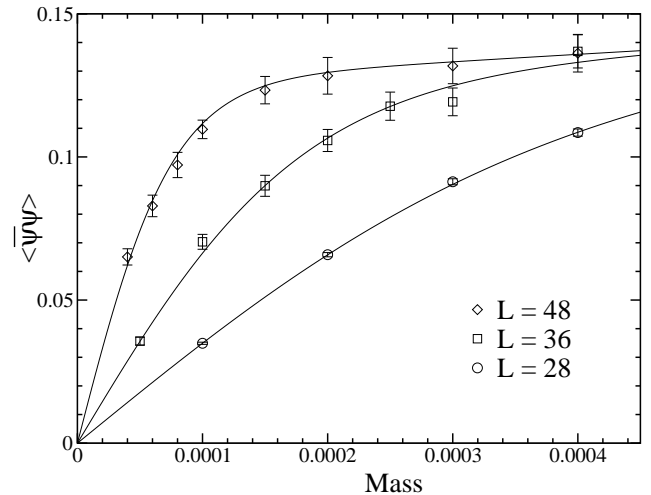}
\end{center}
\vskip-0.5in
\caption{Chiral condensate as a function of the mass at three different 
lattice sizes at $T=1.0471$. The data from the simulations is plotted 
along with the function (\ref{fss1}) for each volume. The same values 
of $\Sigma_0$ and $\chi_0$ are used for all curves shown.}
\label{tanh}
\end{figure}

The above finite size scaling analysis works exceptionally well,
as can be seen from figure \ref{tanh}, which shows a plot of the
condensate as a function of the mass at a fixed $T=1.0471$ for 
three different lattice sizes. The solid lines represent the function 
(\ref{fss1}) for a fixed value of $\Sigma_0$ and $\chi_0$ obtained 
from a single fit to all of the shown data and more. The $\chi^2/DOF$ 
for the fit is around 0.8. Since $T=1.0471$ is very close to the 
critical temperature it was necessary to go to spatial volumes 
as large as $48^3$ in order to determine $\Sigma_0$ with reasonable 
precision. 

Interestingly, the above fitting procedure failed to yield an acceptable
chi-squared for data above a certain temperature. This was essentially 
due to the fact that it was impossible to find a small enough mass to 
ensure that the condition (\ref{mcond}) was satisfied. Removing the larger 
mass data resulted in a smaller $\Sigma_0$ which in turn lowered the bound 
in (\ref{mcond}). This observation is consistent with a vanishing condensate 
in the chiral limit for higher temperatures.

\begin{figure}
\begin{center}
\includegraphics[width=0.45\textwidth]{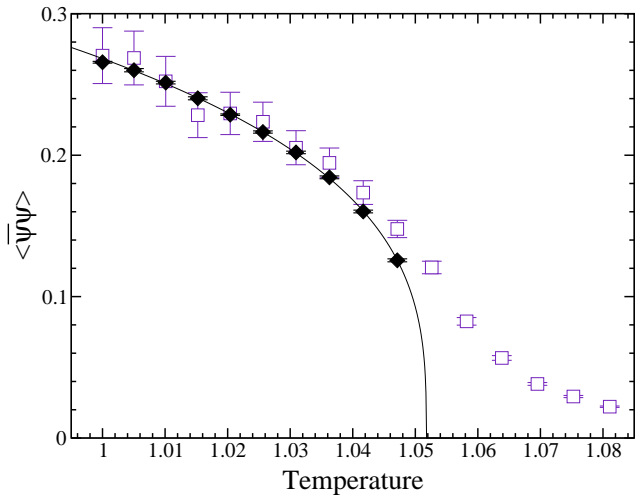}
\end{center}
\vskip-0.5in
\caption{ Chiral condensate versus temperature.
The open squares are results for $L=32$ and $m=0.001$.
The filled diamonds are the extrapolation to the infinite volume limit and
the chiral limit. The solid line is a plot of $A(1.0518-T)^{0.314}$ obtained 
from a fit of the extrapolated points.}
\label{ccm0}
\end{figure}
The non-zero values of $\Sigma_0$ extracted from the fits at different 
temperatures are shown in figure \ref{ccm0}. The solid line represents a 
fit to the expected form $A (T_c-T)^\beta$ using the data for 
temperatures between 1.0050 and 1.0471. The $\chi^2/DOF$ for the fit
was 0.4. The critical exponent $\beta$ was found to be $0.314(7)$ 
which is consistent with the measured value of 0.32648(18) in the Ising 
model \cite{Ising} within two sigma. The fit also determines the critical 
temperature 
very accurately to be $T_c = 1.0518(3)$ which agrees with the 
previously obtained result \cite{Cha99.3} within errors. The data
for $L=32$ and $m=0.001$ is shown for comparison and to demonstrate the
necessity of using the finite size scaling formula (\ref{fss1})
to extract the infinite volume chiral limit. Furthermore, the errors on
$\Sigma_0$ are typically much smaller than those for any single data
point due to the constraints that arise from fitting over a wide range
of masses and volumes.

At $T_c$ the finite size scaling behavior of the condensate as a 
function of the mass is very different from (\ref{fss1}) and can
be shown to be of the form
\begin{equation}
\langle \bar\psi\psi \rangle = L^{y_m-d} f(L^{y_m} m)
\label{fssm}
\end{equation}
using renormalization group arguments that are valid close to a second 
order phase transition. For any $\mathbf{Z}_2$ phase transition the function 
$f(x)$ is universal and depends only on the lattice geometry, boundary 
conditions etc., and not on the type of lattice, irrelevant operators 
and such. The exponent $y_m$ is hence universal. The behavior of $f(x)$ 
is known in the two limits:
\begin{eqnarray}
f(x) &\stackrel{x \rightarrow 0}{\longrightarrow}& f_0\;x \;,
\label{fssm1}
\\
f(x) &\stackrel{x \rightarrow \infty}{\longrightarrow}& 
 f_\infty \; x^{\frac{1}{\delta}} \;,
\label{fssm2}
\end{eqnarray}
where $1/\delta = (d-y_m)/y_m$.

\begin{figure}
\begin{center}
\includegraphics[width=0.45\textwidth]{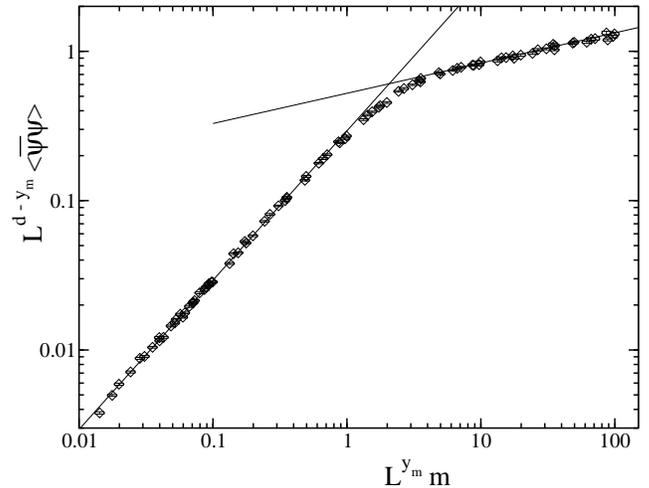}
\end{center}
\vskip-0.5in
\caption{Log-log plot of $L^{d-y_m}\;\langle \bar\psi\psi\rangle$
versus $x \equiv L^{y_m}m$ which describes the scaling function
given in (\ref{fssm}). The value of $y_m$ used is obtained from a 
fit to the small $x$ data as explained in the text.
The solid lines are fits to (\ref{fssm1}) for small $x$ 
and (\ref{fssm2}) for large $x$ data.}
\label{ccs}
\end{figure}
Based on the estimate of $T_c$ obtained above, additional simulations 
at $T=1.0518$ were performed to confirm this critical behavior as a 
function of the mass. Using the data with $L \ge 8$ and 
$L^{y_m} m \le 0.1$, the chiral condensate is fit to the form
\begin{eqnarray}
\langle \bar\psi\psi \rangle \approx f_0 L^{2 y_m - d} m
\label{fss2}
\end{eqnarray}
based on the expectation from (\ref{fssm1}).
Using this value of $y_m$, we can calculate
\begin{equation}
\delta = 4.87(10)
\end{equation}
which compares favorably with the known value for the Ising model
of 4.7893(22) \cite{Ising}. Using the fitted value of $y_m$, the quantity 
$L^{d-y_m} \langle \bar\psi\psi \rangle$ is plotted in figure \ref{ccs}
for all available data as a function of $L^{y_m} m$.
Clearly the data collapses onto a single curve within errors,
as expected from (\ref{fssm}) . To confirm this 
picture the data in figure \ref{ccs} with $L \ge 12$ and
$10<L^{y_m} m<90$ is fit to the form (\ref{fssm2}).
This yields the value $\delta = 4.89(19)$ 
which agrees with the earlier determination and 
further demonstrates that the critical behavior in the present model 
is consistent with the 3-d Ising universality class.

\section{Conclusions}

The meron cluster algorithm has allowed a direct simulation of
a strongly interacting fermionic theory in the vicinity of a $\mathbf{Z}_2$ 
chiral phase transition on lattices with spatial volumes of up to $48^3$
using common workstation computers. The scaling behavior near the critical 
point, as a function of both the temperature and mass, was determined 
within errors of about $2\%$ and the results are consistent with a 
second order phase transition. Allowing for 1-2 sigma deviations
the universality class of the transition matches with that of the
3-d Ising model. This result strongly supports the scenario where
a fermionic theory in $d+1$ dimensions undergoes a dimensional reduction
to be described by a $d$ dimensional bosonic theory in the critical region.
Comparing similar results with the hybrid Monte Carlo algorithm, 
the meron algorithm provides great improvement over such methods.

\vskip0.5in

{\bf Acknowledgement}
\medskip

We would like to thank Uwe Wiese for many useful discussions.
This work was supported in part by a grant from the US Department 
of Energy, Office of Energy Research (DE-FG02-96ER40945). The
computations were performed on {\bf Brahma}, a Pentium based 
Beowulf cluster constructed using computers donated generously by 
the Intel Corporation and located in the physics department at Duke 
University.

\end{document}